\documentclass[aps,prb,twocolumn,superscriptaddress,showpacs]{revtex4-1}

\usepackage{graphicx}
\usepackage{calc}
\usepackage{bm}
\usepackage{color}
\usepackage{textcomp}

\begin{document}

\title{Spin-polarized electron transport for the altermagnet CrSb}

\author{V.D.~Esin}
\author{D.Yu.~Kazmin}
\author{A.V.~Timonina}
\author{N.N.~Kolesnikov}
\affiliation{Institute of Solid State Physics of the Russian Academy of Sciences, Chernogolovka, Moscow District, 2 Academician Ossipyan str., 142432 Russia}
\author{E.V.~Deviatov}
\affiliation{Institute of Solid State Physics of the Russian Academy of Sciences, Chernogolovka, Moscow District, 2 Academician Ossipyan str., 142432 Russia}
\affiliation{V.L. Ginzburg Research Centre for High-Temperature Superconductivity and Quantum Materials, P.N. Lebedev Physical Institute of RAS, Moscow 119991, Russia}

\date{\today}

\begin{abstract}

We experimentally investigate spin-polarized electron transport for the centrosymmetric altermagnet CrSb, which is known to reveal both altermagnetic and topological features. We demonstrate pronounced first-harmonic anomalous and second-harmonic non-linear Hall effects for a single-crystal  CrSb flake with ferromagnetic nickel contacts, while both effects can not be seen for the reference samples with non-magnetic gold ones.
For the anomalous Hall effect, we demonstrate bow-tie  hysteresis loop in Hall voltage, which is usually ascribed to surface spin textures in magnetic materials. The slope of the Hall curve  changes a sign for two orientations of the Hall-bar contact configuration for the same sample, i.e. for the same sign of the charge carriers. We interpret the observed sign inversion and bow-tie hysteresis as the joint effect of  the alternating bulk spin splitting and spin-polarized topological surface states in CrSb. The pronounced  non-linear Hall effect with hysteresis in magnetic field confirms finite Berry curvature dipole under injection of spin-polarized electrons, i.e. the topological features for the  altermagnetic candidate CrSb.

\end{abstract}

\pacs{73.40.Qv  71.30.+h}

\maketitle

\section{Introduction}

Recently, altermagnetic class of materials has been introduced~\cite{alter_common1, alter_common2, alter_mazin} as antiferromagnets with broken  the combined spatial-inversion and time-reversal symmetry~\cite{alter_ferro}. In this case, the spin-opposite sublattices cannot be related by translation or inversion~\cite{alter1,alter2}, so the up-polarized subband is obtained by $\pi/2$ rotation of the down-polarized one for the simplest example of the d-wave altermagnet~\cite{alter_supercond_notes,alter_normal_junction}. As a result, the zero net magnetization is accompanied by alternating spin splitting in the k-space~\cite{alter_common1,alter1,alter_josephson}. 

For altermagnets, the anomalous Hall effect (AHE) as a hysteresis loop in Hall resistance has been regarded as the first experimental demonstration of their non-trivial spin splitting~\cite{AHE_RuO2,AHE_MnTe1,AHE_MnTe2,AHE_Mn5Si3,Mn5Si3_1}. While weak ferromagnetism of antiferromagnets has already been known for materials with strong spin-orbit coupling~\cite{weak_dz}, the specifics of altermagnetism is the non-relativistic spin-momentum locking~\cite{alter_common1}. 

Usual spin-orbit induced spin-momentum locking is well known for topological materials, leading to spin polarization of the topological surface states~\cite{Volkov-Pankratov,MZHasan,Armitage,PhysRevB.100.195134}. In magnetic topological semimetals, the anomalous Hall effect is connected with dissipationless surface transport~\cite{Armitage}. For non-magnetic systems, AHE is prohibited in the presence of time-reversal symmetry, while the non-linear Hall effect (NLHE)~\cite{sodemann} is  predicted as a transverse second-harmonic voltage response due to the Berry curvature dipole in momentum space~\cite{sodemann,deyo,golub,moore,low}. In the simplified picture~\cite{sodemann}, the longitudinal current generates the effective sample magnetization, which leads to the Hall effect even in zero external magnetic field. Hall voltage is therefore  proportional to the square of the excitation current, so it can be detected as the second-harmonic  $V^{2\omega}_{xy}$ transverse voltage component.  NLHE has been demonstrated for a wide range of materials~\cite{ma,kang,esin,c_axis,gete2w}, being the  direct manifestation of finite Berry curvature in topological media~\cite{sodemann}.   

Nowadays, it is a common  agreement, that AHE still requires  spin-orbit coupling  in altermagnetic  materials~\cite{AHE_MnTe1,AHE_MnTe2,spin_ferro_soc,orbital_mag1,SO_AHE_26}. For example, ARPES (angular-resolved photoemission spectroscopy) confirms the altermagnetic nature of spin-spliting in $\alpha$-MnTe~\cite{MnTe_ARPES2}. However, the principle origin of finite net  magnetization~\cite{AHE_MnTe1,AHE_MnTe2,orlova_MnTe1,orlova_MnTe2} in MnTe is the spin-orbit coupling~\cite{satoru,MnTe_SO,orbital_mag1,SO_AHE_26,Dichroism}. In contrast, spin-orbit coupling is negligible  for centrosymmetric CrSb, while ARPES confirms high altermagnetic splitting for this material~\cite{ARPES1_CrSb, ARPES2_CrSb}. Moreover, signature of topological surface states~\cite{AMtopology1,AMtopology2} has been shown for CrSb~\cite{Weyl alter1_CrSb}, so the room-temperature altermagnetic candidate CrSb reveals both altermagnetic and topological features~\cite{Weyl alter2_CrSb}. 

AHE can not be expected  for CrSb either from the topological surface states due to the time-reversal symmetry~\cite{Armitage} or from the bulk altermagnetic spectrum without spin-orbit coupling~\cite{orbital_mag1,SO_AHE_26}. Similarly, NLHE is prohibited for  centrosymmetric materials~\cite{sodemann,deyo,golub,moore,low}. However, both AHE and NLHE becomes more intriguing in spin-injection experiments~\cite{mizuno} due to the alternating bulk spin splitting~\cite{ARPES1_CrSb, ARPES2_CrSb} and strong spin polarization of the topological surface states~\cite{Weyl alter1_CrSb,Weyl alter2_CrSb}.

Here, we experimentally investigate spin-polarized electron transport for the centrosymmetric altermagnet CrSb, which is known to reveal both altermagnetic and topological features. We demonstrate pronounced anomalous and second-harmonic non-linear Hall effects for a single-crystal  CrSb flake with ferromagnetic nickel contacts, while both effects can not be seen for the reference samples with non-magnetic gold ones.

\section{Samples and technique}

\begin{figure}
\includegraphics[width=0.8\columnwidth]{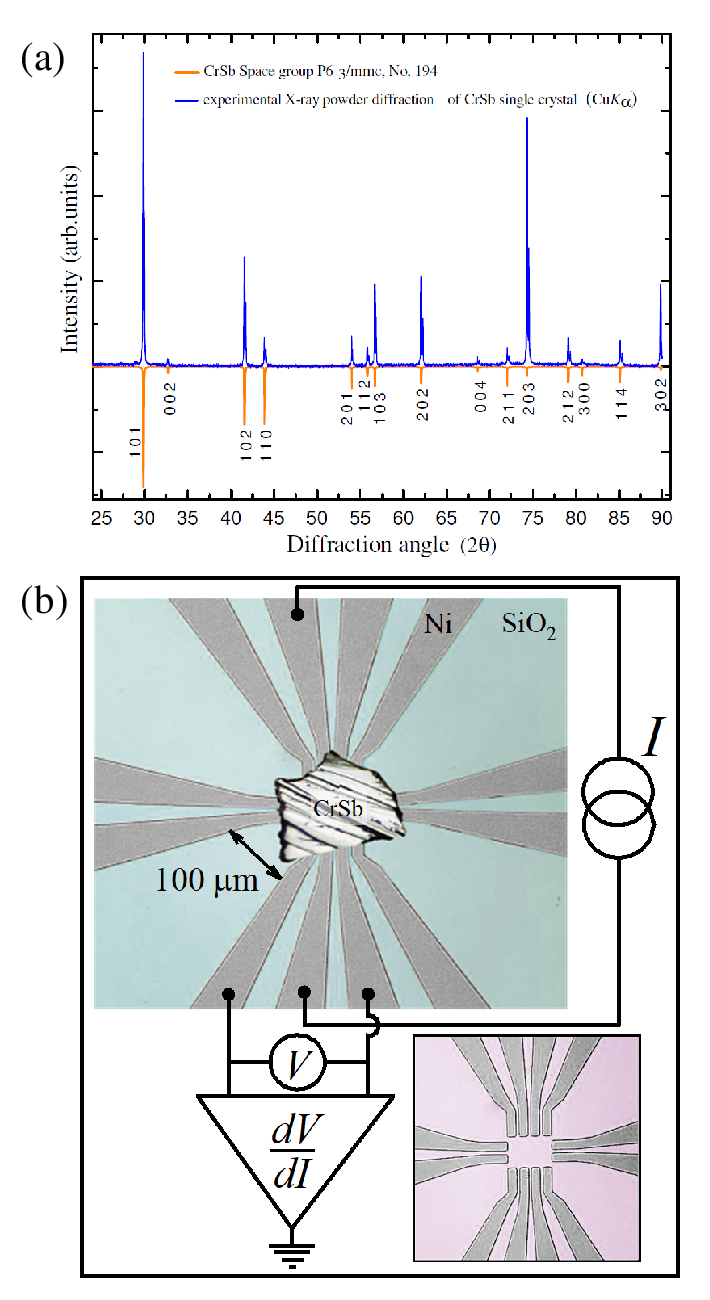}
\caption{(Color online) (a) The X-ray powder diffraction  pattern (Cu K$_{\alpha}$ radiation), which is obtained for the crushed CrSb single crystal. The single-phase  CrSb is confirmed with the space group $P6_3 /mmc$ (No. 194).
(b) Optical image of the sample with a 1~$\mu$m thick CrSb flake on the ferromagnetic Ni leads with scheme of electrical connections for one of two Hall configurations. The leads pattern is shown in the inset: 100 nm thick, 5~$\mu\mbox{m}$ separated leads allow to realize the correct Hall (xy-) configuration for two  orientations in respect to the CrSb flake.  The depicted Hall configuration is specially designed to exclude possible admixture of the thermoelectric effects for reliable NLHE measurements~\cite{esin}.
  }
\label{fig1}
\end{figure}

CrSb single crystals were synthesized by reaction of pure Cr (99.996\%) and Sb (99.9999\%), which were mixed in the stoichiometric ratio and then heated in an evacuated silica ampule up to 1000$^\circ$C with the rate of 15$^\circ$C/h in a gradient-free furnace. The load was held at 1000$^\circ$C for 72 hours and then cooled down slowly (11$^\circ$C/h) to the room temperature. The crystals grown are faceted single crystals with the space group $P6_3 /mmc$ (No. 194) and the stoichiometric composition, as confirmed by X-ray diffraction analysis, see Fig.~\ref{fig1} (a). 

CrSb is of hexagonal structure with the magnetic space group  $P6'_3/m'm'c$: two Cr sublattices with opposite spins are aligned along the $c$ axis, being connected by $C6z$  -  a six-fold rotation  combined with  a 1/2 translation along c~\cite{neudiff1, neudiff2}. The N\'eel temperature is about 705~K, which might be convenient for possible applications in spintronics. 

Fig.~\ref{fig1} (b) shows a top-view image of the sample.  Since CrSb  is a three-dimensional altermagnet, we have to select  relatively thick (about 1~$\mu$m) single-crystal flakes. A thick flake  also ensures sample homogeneity for correct determination of xx- and xy- voltage responses, however, the desired experimental geometry can not be defined by usual mesa etching for thick flakes. Instead, the contact geometry is formed by Ni leads pattern on the SiO$_2$ surface, as depicted in the inset to Fig.~\ref{fig1} (b). Ni leads are obtained by lift-off technique after thermal evaporation of 100~nm Ni film.  As a reference, we use Au leads pattern for the samples without spin polarization of charge carriers. 

The mechanically exfoliated $\approx$100~$\mu$m wide CrSb flake is transferred onto the metallic  leads. After initial single-shot pressing by another oxidized silicon substrate,  the flake is firmly connected to the leads, see Fig.~\ref{fig1} (b). This procedure provides high-quality contacts, the obtained samples are electrically and mechanically  stable even in different cooling cycles, which has been verified  before for a wide range of materials~\cite{crsbin,mntein,cosns,timnal,black}. Also,  the Ni-CrSb (or Au-CrSb) contacts are protected from any contamination (also oxygen  and  moisture) by SiO$_2$ substrate, as it has been demonstrated for sensitive materials like black phosphorus~\cite{black}. 

We investigate  transverse (xy-)  first- and second-harmonic  voltage responses by standard four-point lock-in technique~\cite{esin,gete2w}. The ac current is applied between two current contacts in Fig.~\ref{fig1} (b), while the transverse (Hall) voltage is measured between the  potential ones.  The signal is confirmed to be independent of the ac current frequency within 100 Hz -- 10kHz range, which is defined by the applied filters.
The measurement are performed at 30 mK temperature in a dilution refrigerator equipped with a superconducting solenoid. Magnetic field is directed normally to the sample plane to investigate the first- and second-harmonic Hall effects. 

The leads pattern in the inset to Fig.~\ref{fig1} (b) allows to realize the correct Hall (xy-) configuration for two different Hall bar orientations in respect to the CrSb flake: one configuration~\cite{esin} is shown in Fig.~\ref{fig1} (b), while the current line is 90$^\circ$ rotated for the other one~\cite{gete2w}. In the contacts area, the leads are of 10~$\mu$m width, they are separated by 5~$\mu$m distance, so the voltage probe separation is 20-80~$\mu$m for two different Hall configurations. The correct choice of the potential contacts is important for AHE and NLHE measurements. In particular, the  contacts configuration from Fig.~\ref{fig1} (b) is specially designed for reliable NLHE measurements, it has been verified to exclude possible admixture of the thermoelectric effects, see Ref.~\cite{esin} for details.

\section{Experimental results}

\begin{figure}
\includegraphics[width=\columnwidth]{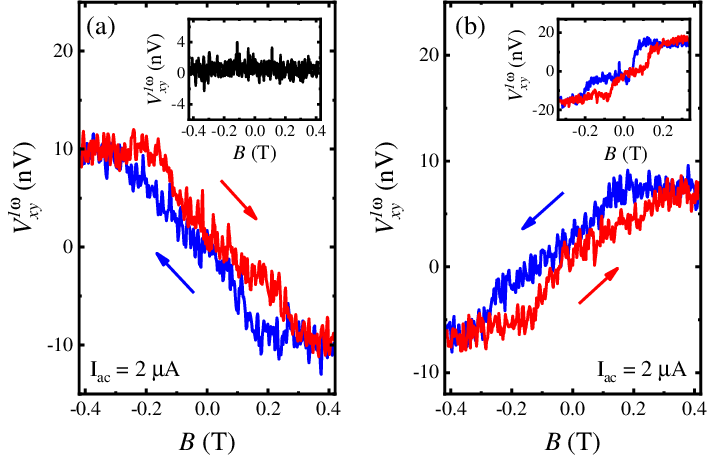}
\caption{(Color online)  Anomalous Hall effect for the CrSb sample with ferromagnetic Ni contacts. (a) AHE hysteresis is depicted as the magnetic field dependence of the transverse voltage $V^{1\omega}_{xy}$ for two field sweep directions for the contacts configuration from Fig.~\ref{fig1} (b).  Inset shows $V^{1\omega}_{xy}(B)$ curves for the reference sample with normal gold contacts, no AHE can be seen in this case. (b) $V^{1\omega}_{xy}(B)$ hysteresis for another Hall configuration, which is 90$^\circ$ rotated in respect to the one in (a). The AHE slope is inverted in this case for the same CrSb flake and for the same magnetic field orientation. For both Hall configurations, $V^{1\omega}_{xy}(B)$ curves strongly resemble bow-tie magnetization hysteresis loops~\cite{bow-tie,cosnsmag}, which is confirmed in the inset for another sample. The data are obtained at $T=30$~mK temperature for the ac current amplitude $I_{ac}=2 \mu$A. 
 }
\label{fig2}
\end{figure}

Fig.~\ref{fig2} (a) shows typical AHE signal  as the magnetic field dependence of the transverse voltage $V^{1\omega}_{xy}$  for the sample with ferromagnetic Ni contacts. The data are obtained for the contacts configuration from Fig.~\ref{fig1} (b) at $T=30$~mK temperature. The ac current amplitude $I_{ac}=2 \mu$A guarantees the linear regime, e.g. it do not lead to the sample overheating for the low characteristic longitudinal (xx-) sample resistance $\approx 0.5 \Omega$. The maximum Hall resistance is 5~m$\Omega$ in Fig.~\ref{fig2} (a), it corresponds to Hall resistivity 0.5~$\mu\Omega\cdot$cm for 1~$\mu$m thick CrSb flake, which is at least one order of magnitude smaller than the usual values in magnetic materials~\cite{AHE_MnTe1,AHE_MnTe2,planarHall}.  To improve the signal to noise ratio for low $V^{1\omega}_{xy}(B)$ values, we use multiple curves averaging, so the hysteresis in $V^{1\omega}_{xy}(B)$ can be clearly seen for two magnetic field sweep directions in Fig.~\ref{fig2} (a). 

This result is quite unusual for  CrSb, since no AHE can be expected without spin-orbit coupling in altermagnetic  materials~\cite{satoru,MnTe_SO,orbital_mag1,SO_AHE_26}. We also do not observe any noticeable AHE signal  for the reference samples with normal gold contacts, as depicted in the inset to Fig.~\ref{fig2} (a), so the observation of AHE should be a consequence of the spin-polarized current injection from the ferromagnetic  Ni contacts.

The leads pattern in the inset to Fig.~\ref{fig1} allows to realize another Hall (xy-) contact configuration, where the current line is 90$^\circ$ rotated in respect to the previous case, the result is depicted in Fig.~\ref{fig2} (b). The $V^{1\omega}_{xy}(B)$ hysteresis is of the same $\pm 0.3$~T width in Fig.~\ref{fig2} (b), but of lower amplitude in comparison with one in Fig.~\ref{fig2} (a), the latter can be ascribed to geometrical factor. However, the slope changes a sign  in Fig.~\ref{fig2} (b), which can not be expected for the same flake and the same magnetic field orientation~\cite{AHE_Mn5Si3,Mn5Si3_1,planarHall}: in normal magnetic field, the angle between the magnetic field and the current is the same for two Hall bar configurations in Fig.~\ref{fig2}. Since the slope of the $V^{1\omega}_{xy}(B)$ Hall curve is determined by the sign of the charge carriers and the magnetic field direction, it should not depend on the current line orientation in respect to the CrSb flake.    

Also, the hysteresis loops strongly resemble a bow-tie magnetization hysteresis loop~\cite{bow-tie,cosnsmag} in Fig.~\ref{fig2}: the distance between the curves for opposite sweep directions is smaller at zero field/zero voltage, which  can be even better seen for another sample in the inset to Fig.~\ref{fig2} (b). The bow-tie hysteresis in magnetization  is usually ascribed to surface spin textures in magnetic materials~\cite{bow-tie,cosnsmag}.

\begin{figure}
\includegraphics[width=0.85\columnwidth]{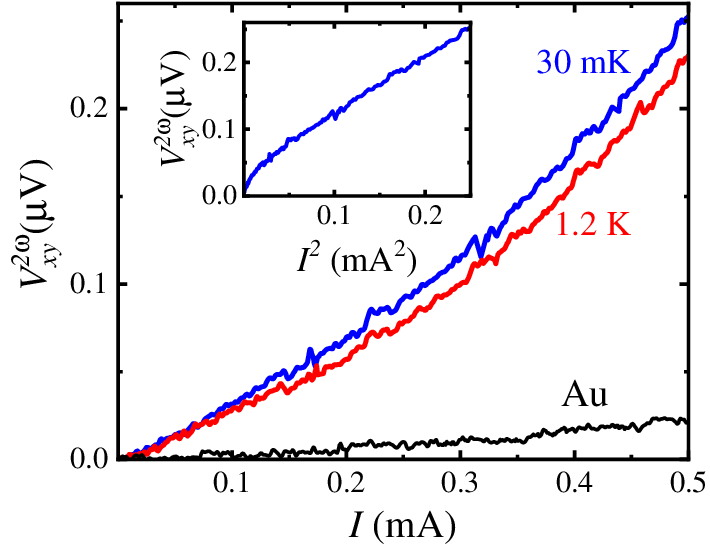}
\caption{(Color online) Non-linear Hall effect~\cite{ma,kang,esin,c_axis,gete2w} for the ferromagnetic Ni contacts as a quadratic second-harmonic Hall-like response $V^{2\omega}_{xy}\sim I^2$ to ac excitation current $I$.  The  curves are depicted for two 30~mK and 1.2~K temperatures as blue and red ones, respectively. $V^{2\omega}_{xy}\sim I^2$ law is confirmed in the inset  as strictly linear $V^{2\omega}_{xy}(I^2)$ dependence for high ac excitation values. The second-harmonic Hall voltage $V^{2\omega}_{xy}$ is one order of magnitude smaller for the reference sample with Au contacts for the same contact configuration, as depicted by the black curve. The data are obtained in zero magnetic field.
}
\label{fig3}
\end{figure}

Fig.~\ref{fig3} shows the effect of spin-polarized current on the NLHE, i.e. on the second harmonic of the transverse (Hall) voltage $V^{2\omega}_{xy}$. For the CrSb sample with ferromagnetic Ni contacts, typical behavior of the non-linear Hall effect~\cite{ma,kang,esin,c_axis,gete2w} is demonstrated as a quadratic transverse Hall-like response $V^{2\omega}_{xy}\sim I^2$ to ac excitation current $I$. In contrast, the second-harmonic Hall voltage is one order of magnitude smaller for the reference sample with Au contacts in   Fig.~\ref{fig3}, the data are obtained for the same contact configuration. For NLHE measurements, we only use the contact configuration depicted in  Fig.~\ref{fig1} (b), to reliably exclude the thermoelectric effects, as it has been demonstrated in Ref.~\cite{esin}.

For NLHE, wide current interval is important to observe non-linear effects. The NHLE curves $V^{2\omega}_{xy}(I)$ still depend on temperature in 30~mK--1.2~K range, see red and blue curves, so the ac excitation do not overheat the sample. Moreover, sample overheating would lead to diminishing of the measured voltage, while the $V^{2\omega}_{xy}\sim I^2$ dependence is well confirmed for high ac excitations  in the inset to Fig.~\ref{fig3}.     

\begin{figure}
\includegraphics[width=\columnwidth]{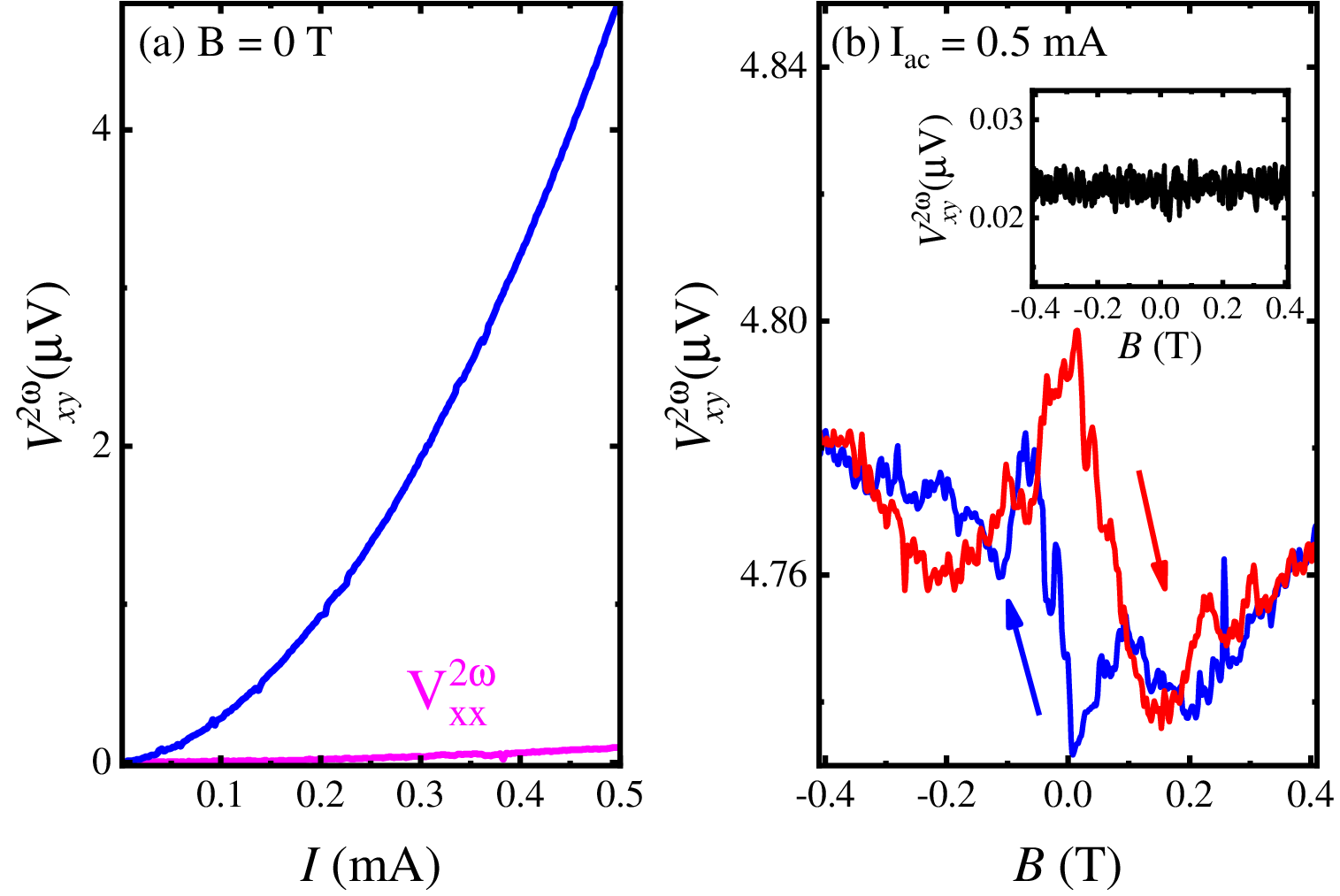}
\caption{(Color online) (a) Non-linear Hall effect for another sample with ferromagnetic Ni contacts in zero magnetic field.  The magenta curve shows the longitudinal  second-harmonic voltage $V^{2\omega}_{xx}$, which is one order of magnitude smaller. (b) Hysteresis in $V^{2\omega}_{xy}$ for two magnetic field sweep directions at $I_{ac}=0.5$~mA for spin-polarized transport.  Inset shows  no $V^{2\omega}_{xy}(B)$ field dependence for the reference sample with gold contacts. The data are obtained at $T=30$~mK temperature. 
}
\label{fig4}
\end{figure}

Qualitatively similar $\sim I^2$  NLHE dependence  is reproduced for another sample with ferromagnetic Ni contacts, see Fig.~\ref{fig4} (a). The longitudinal  second-harmonic voltage $V^{2\omega}_{xx}$ is negligible in comparison with the Hall $V^{2\omega}_{xy}$ one, which confirms well-defined experimental geometry and high sample homogeneity. 

As well as the first harmonic in Fig.~\ref{fig2}, the second-harmonic Hall voltage shows hysteresis for two magnetic field sweep directions, see Fig.~\ref{fig4} (b). The hysteresis loops are of similar widths  in Figs.~\ref{fig2} and.~\ref{fig4} (b).  In contrast, there is no $V^{2\omega}_{xy}(B)$ field dependence for the reference sample with gold contacts, as it is shown in the inset to Fig.~\ref{fig4} (b).  

We should conclude, that for the altermagnet CrSb both the first-harmonic anomalous and the second-harmonic non-linear Hall effects can only be observable for spin-polarized charge carriers.

\section{Discussion}

Both AHE and NLHE  can not be expected  for non-magnetic and  centrosymmetric CrSb, which we confirm for the reference CrSb samples with gold contacts in Figs.~\ref{fig2},\ref{fig3},\ref{fig4}. Also, the stoichiometric CrSb composition was verified  by X-ray diffraction analysis, see Fig.~\ref{fig1} (a), so one can be sure in quality of our CrSb material. Moreover, the AHE slope inversion in Fig.~\ref{fig2} can not be expected even in magnetic materials for the same flake (i.e. for the same sign of the charge carriers) and the same magnetic field orientation. 
Thus, observation of AHE and NLHE should be connected with spin injection into the altermagnetic crystal CrSb. 

Let us start from the anomalous Hall effect in  Fig.~\ref{fig2}. In principle, CrSb exhibits low magnetocrystalline anisotropy energy, enabling the manipulation of the Néel vector in CrSb films through a suitable ferromagnetic substrate, so the anomalous Hall and Nernst conductivities appear for  the canted Néel vector~\cite{ahe5}. However, this effect can only be observable within the bulk spin-relaxation length, i.e. near the contact regions in our experiment. Instead, we demonstrate the anomalous Hall effect for macroscopic (20-80~$\mu$m) contact separation. 

On the other hand, Weyl topological surface states were shown for CrSb~\cite{Weyl alter1_CrSb,Weyl alter2_CrSb}. The states with opposite spin projections are counter-propagating for the vanishing net magnetization in CrSb, so there is no net AHE for the current injection from spin-unpolarized gold contacts~\cite{Armitage}, as we corfirm for the reference samples. In contrast, spin injection from the ferromagnetic Ni contacts leads to the charge imbalance between counter-propagating surface states and, therefore, to the anomalous Hall effect in Fig.~\ref{fig2}. The effect is observable for macroscopic distances due to the topological protection of charge transport within Weyl surface states. The hysteresis width in Fig.~\ref{fig2}  should be connected with Ni contact  remagnetization, i.e., with the change of the spin polarization of the injected carriers. 

However, this picture is only valid for the directions of full spin compensation in bulk altermagnetic spectrum. For other directions, AHE is affected by periodic bulk altermagnetic magnetization~\cite{crsb_magnetization}, which leads to the direction-dependent sign of the AHE slope in Fig.~\ref{fig2}. In this case, the bow-tie hysteresis loops in   Fig.~\ref{fig2} is the result of interaction~\cite{bow-tie,cosnsmag} between the  bulk alternating spin splitting and the surface spin polarization from the topological surface states~\cite{Weyl alter2_CrSb,Weyl alter1_CrSb}. 
Thus, the AHE can be explained as the joint effect of  the k-dependent bulk altermagnetic spin polarization and spin-polarized topological surface states for the spin-polarized charge carriers in the  altermagnetic candidate CrSb.

Similar mechanism of spin injection is valid for the non-linear Hall effect in Figs.~\ref{fig3} and~\ref{fig4}.  In general, NLHE~\cite{ma,kang,esin,c_axis,gete2w} arises from  the Berry curvature dipole in momentum space~\cite{sodemann}. One can not expect non-zero Berry curvature dipole for centrosymmetric CrSb, which is confirmed by negligible $V^{2\omega}_{xy}$ for the reference sample with gold contacts in Fig.~\ref{fig3}. However, spin injection leads to the charge imbalance between the topological surface states, and, in order, to  finite Berry curvature dipole due to the bulk-boundary correspondence, which appears as finite $V^{2\omega}_{xy}$ transverse voltage in Figs.~\ref{fig3} and~\ref{fig4}. 

Another possibility to have finite second-harmonic  response is proposed in Ref.~\cite{golub_alter}. The flow of electric current in an altermagnet could result in the formation of  homogeneous electron spin orientation in the sample, which is quadratic in the current magnitude and does not require broken inversion symmetry. In principle, the  electron spin orientation could be detected by ferromagnetic contacts, however, the effect should be observed primary for the xx- contact configuration. In contrast, the longitudinal  second-harmonic voltage $V^{2\omega}_{xx}$ is negligible in comparison with the Hall $V^{2\omega}_{xy}$ one in Fig.~\ref{fig4} (a). 

The second-order $V^{2\omega}_{xy}(B)$ NLHE response is sensitive to the external magnetic field~\cite{2wMagF1,2wMagF2}. Being determined by spin injection from the ferromagnetic Ni contact, NLHE voltage  shows hysteresis within the same  field range in Fig.~\ref{fig4} (b), as the first-harmonic AHE  response in Fig.~\ref{fig2}. It is worth mentioning, that there is no field dependence for the reference sample, as it is shown in the inset to  Fig.~\ref{fig4} (b). Also, NLHE should be affected by the bulk alternating spin splitting~\cite{crsb_magnetization} similarly to AHE, which might be responsible for very different $V^{2\omega}_{xy}$ values for samples in Figs.~\ref{fig3} and~\ref{fig4} (a).

\section{Conclusion}

As a conclusion, we experimentally investigate spin-polarized electron transport for the centrosymmetric altermagnet CrSb, which is known to reveal both altermagnetic and topological features. We demonstrate pronounced anomalous Hall and second-harmonic non-linear Hall effects for a thin single-crystal  CrSb flake with ferromagnetic nickel contacts, while they can not be seen for the reference samples with non-magnetic gold ones.
For the anomalous Hall effect, we demonstrate bow-tie  hysteresis in Hall voltage, which is usually ascribed to surface spin textures in magnetic materials. The slope of the Hall curve  changes a sign for two Hall bar orientations for the same sample, i.e. for the same sign of the charge carriers. We interpret the observed sign inversion and bow-tie hysteresis as the joint effect of  the k-dependent bulk altermagnetic spin polarization and spin-polarized topological surface states  for the  altermagnetic candidate CrSb. The pronounced  non-linear Hall effect with hysteresis in magnetic field confirms finite Berry curvature dipole under injection of spin-polarized electrons in altermagnet CrSb.

\acknowledgments
We wish to thank S.S~Khasanov for X-ray sample characterization and Vladimir Zyuzin for valuable discussions.  We gratefully acknowledge financial support  by the  Russian Science Foundation, project RSF-26-12-00001, https://rscf.ru/project/26-12-00001/


\begin{thebibliography}{99}


\bibitem{alter_common1} Libor $\check{S}$mejkal, Jairo Sinova, and Tomas Jungwirth, "Beyond Conventional Ferromagnetism and Antiferromagnetism: A Phase with Nonrelativistic Spin and Crystal Rotation Symmetry", Phys. Rev. X 12, 031042 (2022)  DOI: 10.1103/PhysRevX.12.031042
\bibitem{alter_common2} Libor $\check{S}$mejkal, Jairo Sinova and Tomas Jungwirth, "Emerging Research Landscape of Altermagnetism",  Phys. Rev. X 12, 040501 (2022). DOI: 10.1103/PhysRevX.12.040501
\bibitem{alter_mazin} Igor Mazin, "Altermagnetism—A New Punch Line of Fundamental Magnetism", Phys. Rev. X 12, 040002 (2022); 10.1103/PhysRevX.12.040002
\bibitem{alter_ferro}  Sang-Wook Cheong, Fei-Ting Huang, "Altermagnetism with non-collinear spins", npj Quantum Mater. 9, 13 (2024) 
https://doi.org/10.1038/s41535-024-00626-6  
\bibitem{alter1}  J. Krempasky, L. \'Smejkal, S.W. D'Souza,  M. Hajlaoui, G. Springholz, K. Uhlířová, F. Alarab, P. C. Constantinou, V. Strocov, D. Usanov, W. R. Pudelko, R. González-Hernández, A. Birk Hellenes, Z. Jansa, H. Reichlová, Z. Šobáň, R. D. Gonzalez Betancourt, P. Wadley, J. Sinova, D. Kriegner, J. Minár, J. H. Dil and T. Jungwirth,  "Altermagnetic lifting of Kramers spin degeneracy", Nature 626, 517 (2024) https://doi.org/10.1038/s41586-023-06907-7  
\bibitem{alter2}  R. Jaeschke-Ubiergo, V. K. Bharadwaj, T. Jungwirth, L. Šmejkal, and J. Sinova, "Supercell altermagnets", Phys. Rev. B 109, 094425 (2024)  https://doi.org/10.1103/PhysRevB.109.094425
\bibitem{alter_supercond_notes} Igor I. Mazin "Notes on altermagnetism and superconductivity" arxiv:2203.05000
\bibitem{alter_normal_junction} Sachchidanand Das, Dhavala Suri, Abhiram Soori, "Transport across junctions of altermagnets with normal metals and ferromagnets" J. Phys. : Condens. Matter 35, 435302 (2023), https://doi.org/10.1088/1361-648X/acea12
\bibitem{alter_josephson} Jabir Ali Ouassou, Arne Brataas, Jacob Linder, "dc Josephson Effect in Altermagnets", Physical Review Letters 131, 076003 (2023); https://doi.org/10.1103/PhysRevLett.131.076003

\bibitem{AHE_RuO2} Z. Feng, X. Zhou, L. \'Smejkal, L. Wu, Z. Zhu, H. Guo, R. Gonz\'alez-Hern\'andez, X. Wang, H. Yan, P. Qin, X. Zhang, H. Wu, H. Chen, Z. Xia, C. Jiang, M. Coey, J. Sinova, T. Jungwirth, and Z. Liu,"An anomalous Hall effect in altermagnetic ruthenium dioxide", Nat. Electron. 5, 735 (2022).
\bibitem{AHE_MnTe1} R.D. Gonzalez Betancourt, J. Zubac, R. Gonzalez-Hernandez, K. Geishendorf, Z. Soban, G. Springholz, K. Olejnik, L. Smejkal, J. Sinova, T. Jungwirth, S.T.B. Goennenwein, A. Thomas, H. Reichlova, J. Zelezny, and D. Kriegner, "Spontaneous Anomalous Hall Effect Arising from an Unconventional Compensated Magnetic Phase in a Semiconductor" Phys. Rev. Lett. 130, 036702 (2023).
\bibitem{AHE_MnTe2}  K. P. Kluczyk, K. Gas, M. J. Grzybowski, P. Skupi\'nski, M. A. Borysiewicz, T. Fas, J. Suffczy\'nski, J. Z. Domagala, K. Grasza, A. Mycielski, M. Baj, K. H. Ahn, K. V\'yborn\'y, M. Sawicki, M. Gryglas-Borysiewicz, "Coexistence of anomalous Hall effect and weak magnetization in a nominally collinear antiferromagnet MnTe" Physical Review B 110, 155201 (2024)
\bibitem{AHE_Mn5Si3} Miina Leivisk\"a, Javier Rial, Antonín Badura, Rafael Lopes Seeger, Ismaila Kounta, Sebastian Beckert, Dominik Kriegner, Isabelle Joumard, Eva Schmoranzerov\'a, Jairo Sinova, Olena Gomonay, Andy Thomas, Sebastian T.B. Goennenwein, Helena Reichlov\'a, Libor \'Smejkal, Lisa Michez, Tom\'a\'s Jungwirth, Vincent Baltz "Anisotropy of the anomalous Hall effect in thin films of the altermagnet candidate Mn5Si3" Phys. Rev. B 109, 224430 (2024)
\bibitem{Mn5Si3_1}   H. Reichlova, R. L. Seeger, R.González-Hernández, I. Kounta, R. Schlitz, D. Kriegner, M. Lammel,  V. Petřiček, Ph. Ritzinger, M. Leiviskä, A. B. Hellenes, P. Doležal, L. Horak, , K. Olejník, E. Schmoranzerova, S. Bertaina, A. Thomas, A. Badura, L. Michez, J. Sinova, S. T. B. Goennenwein, V. Baltz, T. Jungwirth and L. Šmejkal, "Observation of a spontaneous anomalous Hall response in the Mn$_5$Si$_3$ d-wave altermagnet candidate", Nature Communications  Vol 15, 4961 (2024) 

\bibitem{weak_dz} I. Dzyaloshinsky, "A thermodynamic theory of “weak” ferromagnetism of antiferromagnetics" Journal of Physics and Chemistry of Solids 4, 241 (1958).



\bibitem{Volkov-Pankratov} B.A. Volkov and O.A. Pankratov, "Two-dimensional massless electrons in an inverted contact", JETP Letters, 42, 178 (1985) 
\bibitem{MZHasan} M. Z. Hasan and C. L. Kane, "Colloquium: Topological insulators", Reviews of Modern Physics, Vol. 82, pp. 3045--3067 (2010)
\bibitem{Armitage} N. P. Armitage, E. J. Mele, and Ashvin Vishwanath, "Weyl and Dirac semimetals in three-dimensional solids" Rev. Mod. Phys. 90, 15001 (2018)
\bibitem{PhysRevB.100.195134} B. Ghosh, D. Mondal, C.-N. Kuo, C. S. Lue, J. Nayak,
J. Fujii, I. Vobornik, A. Politano, and A. Agarwal, "Observation of bulk states and spin-polarized topological surface states in transition metal dichalcogenide Dirac semimetal candidate NiTe2", Phys. Rev. B 100, 195134 (2019).


\bibitem{sodemann} Inti Sodemann, Liang Fu., "Quantum nonlinear Hall effect induced by Berry curvature dipole in time-reversal invariant materials" Phys. Rev. Lett. 115, 216806 (2015)
\bibitem{deyo} E. Deyo, L. E. Golub, E. L. Ivchenko, and B. Spivak, "Semiclassical theory of the photogalvanic effect in non-centrosymmetric systems" arXiv:0904.1917 (2009).
\bibitem{golub} L.E. Golub, E.L. Ivchenko, B.Z. Spivak, "Photocurrent in gyrotropic Weyl semimetals" JETP Letters, 105, 782 (2017)
\bibitem{moore} J. E. Moore and J. Orenstein, "Confinement-induced Berry phase and helicity-dependent photocurrents" Phys. Rev. Lett., 105, 026805 (2010).
\bibitem{low} T. Low, Y. Jiang, and F. Guinea, "Topological currents in black phosphorus with broken inversion symmetry" Physical Review B 92, 235447 (2015).

\bibitem{ma} Qiong Ma, Su-Yang Xu, Huitao Shen, David MacNeill, Valla Fatemi, Tay-Rong Chang, Andrés M. Mier Valdivia, Sanfeng Wu, Zongzheng Du, Chuang-Han Hsu, Shiang Fang, Quinn D. Gibson, Kenji Watanabe, Takashi Taniguchi, Robert J. Cava, Efthimios Kaxiras, Hai-Zhou Lu, Hsin Lin, Liang Fu, Nuh Gedik and Pablo Jarillo-Herrero, "Observation of the nonlinear Hall effect under time-reversal-symmetric conditions"  Nature 565, 337 (2019).
\bibitem{kang} K. Kang, T. Li, E. Sohn, J. Shan, and K. F. Mak, "Nonlinear anomalous Hall effect in few-layer WTe2"  Nature Mater. 18, 324 (2019).
\bibitem{esin} O. O. Shvetsov, V. D. Esin, A. V. Timonina, N. N. Kolesnikov, and E. V. Deviatov, "Nonlinear Hall effect in three-dimensional Weyl and Dirac semimetals" JETP Letters, 109, 715 (2019). 	DOI: 10.1134/S0021364019110018
\bibitem{c_axis}  A. Tiwari, F. Chen, Sh. Zhong, E. Drueke, J. Koo, A. Kaczmarek, C. Xiao, J. Gao, X. Luo, Q. Niu, Y. Sun, B. Yan, L. Zhao and A. W. Tsen, "Giant c-axis nonlinear anomalous Hall effect in Td-MoTe2 and WTe2" Nat. Commun. 12, 2049 (2021). https://doi.org/10.1038/s41467-021-22343-5
\bibitem{gete2w} N. N. Orlova, A. V. Timonina, N. N. Kolesnikov, and E. V. Deviatov, "Gate-Dependent Nonlinear Hall Effect at Room Temperature in Topological Semimetal GeTe", 	Chinese Physics Letters 40, 077302 (2023) 	https://doi.org/10.1088/0256-307X/40/7/077302



\bibitem{spin_ferro_soc} Merce Roig, Yue Yu, Rune C. Ekman, Andreas Kreisel, Brian M. Andersen, Daniel F. Agterberg, "Quasi-symmetry Constrained Spin Ferromagnetism in Altermagnets"	 Phys. Rev. Lett. 135, 016703 (2025) 
\bibitem{orbital_mag1} Chao Chen Ye, Karma Tenzin, Jagoda Sławińska, Carmine Autieri, "Dominant orbital magnetization in the prototypical altermagnet MnTe",  Phys. Rev. B 113, 014413 (2026)
\bibitem{SO_AHE_26} Yufei Zhao, Yiyang Jiang, Kamal Das, Chao-Xing Liu, Binghai Yan, "Residual orbital magnetization governs the anomalous Hall effect in altermagnets", 	arXiv:2606.25999

\bibitem{MnTe_ARPES2} T. Osumi, S. Souma, T. Aoyama, K. Yamauchi, A. Honma, K. Nakayama, T. Takahashi, K. Ohgushi, and T. Sato,"Observation of Giant Band Splitting in Altermagnetic MnTe",  Phys. Rev. B 109, 115102 (2024)


\bibitem{orlova_MnTe1} N.N. Orlova, A.A. Avakyants, A.V. Timonina, N.N. Kolesnikov, and E.V. Deviatov, "Crossover from relativistic to non-relativistic net magnetization for MnTe altermagnet candidate",  JETP Letters,  120, 360 (2024). 	https://doi.org/10.1134/S0021364024602926
\bibitem{orlova_MnTe2}  N. N. Orlova, V. D. Esin, A. V. Timonina , N. N. Kolesnikov and E. V. Deviatov, "Magnetization symmetry for the altermagnetic candidate MnTe", Phys. Rev. B 111, 224414 (2025) 	DOI: https://doi.org/10.1103/br1r-bjzk
 
\bibitem{satoru} Satoru Hayami and Hiroaki Kusunose, "Essential role of the anisotropic magnetic dipole in the anomalous Hall effect" Phys. Rev. B 103, L180407 (2021). 
\bibitem{MnTe_SO} M. Hajlaoui, S.W. D'Souza, L. Smejkal, D. Kriegner, G. Krizman, T. Zakusylo, N. Olszowska, O. Caha, J. Michalička, A. Marmodoro, K. Výborný, A. Ernst, M. Cinchetti, J. Minar, T. Jungwirth, G. Springholz, "Temperature Dependence of Relativistic Valence Band Splitting Induced by an Altermagnetic Phase Transition"	Adv. Mater.  36, 2314076 (2024)
\bibitem{Dichroism} A. Hariki, A. Dal Din, O. J. Amin, T. Yamaguchi, A. Badura, D. Kriegner, K. W. Edmonds, R. P. Campion, P. Wadley, D. Backes, L. S. I. Veiga, S. S. Dhesi, G. Springholz, L. Smejkal, K. Vyborny, T. Jungwirth, J. Kunes, "X-Ray Magnetic Circular Dichroism in Altermagnetic $\alpha$-MnTe" Phys. Rev. Lett. 132, 176701 (2024)


\bibitem{ARPES1_CrSb} Sonka Reimers, Lukas Odenbreit, Libor Smejkal, Vladimir N. Strocov, Procopios Constantinou, Anna Birk Hellenes, Rodrigo Jaeschke Ubiergo, Warlley H. Campos, Venkata K. Bharadwaj, Atasi Chakraborty, Thiboud Denneulin, Wen Shi, Rafal E. Dunin-Borkowski, Suvadip Das, Mathias Kläui, Jairo Sinova, Martin Jourdan, "Direct observation of altermagnetic band splitting in CrSb thin films", Nat Commun 15, 2116 (2024). https://doi.org/10.1038/s41467-024-46476-5
\bibitem{ARPES2_CrSb} G. Yang, Zh. Li, S. Yang, J. Li, H. Zheng, W. Zhu, Z. Pan, Y. Xu, S. Cao, W. Zhao, A. Jana, J. Zhang, M. Ye, Yu Song, L.-H. Hu, L. Yang, J. Fujii, I. Vobornik, M. Shi, H. Yuan, Y. Zhang, Y. Xu and Y. Liu, "Three-dimensional mapping of the altermagnetic spin splitting in CrSb",  Nature Communications Vol. 16, 1442, pp. 1 (2025)
\bibitem{AMtopology1} Daniil S. Antonenko, Rafael M. Fernandes, Jorn W. F. Venderbos, "Mirror Chern Bands and Weyl Nodal Loops in Altermagnets", Phys. Rev. Lett. 134, 096703 (2025) 
\bibitem{AMtopology2} Rafael M. Fernandes, Vanuildo S. de Carvalho, Turan Birol, Rodrigo G. Pereira, "Topological transition from nodal to nodeless Zeeman splitting in altermagnets", Phys. Rev. B 109, 024404 (2024).
\bibitem{Weyl alter1_CrSb} Cong Li, Mengli Hu, Zhilin Li, Balasubramanian Thiagarajan, Yang Wang, Wanyu Chen, Mats Leandersson, Craig Polley, Cosma Fulga, Maia G. Vergniory, Oleg Janson, Timur Kim, Oscar Tjernberg, Jeroen van den Brink, "Topological Weyl altermagnetism in CrSb",  Communications Physics 8, 311, pp. 1 (2025)
\bibitem{Weyl alter2_CrSb} Wenlong Lu, Shiyu Feng, Yuzhi Wang, Dong Chen, Zihan Lin, Xin Liang, Siyuan Liu, Wanxiang Feng, Kohei Yamagami, Junwei Liu, Claudia Felser, Quansheng Wu and Junzhang Ma, "Signature of Topological Surface Bands in Altermagnetic Weyl Semimetal CrSb", Nano Letters,  Vol 25, 18,  pp. 7343, (2025)

\bibitem{mizuno} Koki Mizuno "High-harmonic spin-current signatures of altermagnetic spin-group symmetry" 	arXiv:2606.31573




\bibitem{neudiff1} A. I. Snow, "Neutron diffraction investigation of the atomic magnetic  moment orientation in the antiferromagnetic compound CrSb", Phys. Rev. 85, 365 (1952)
\bibitem{neudiff2} W. J. Takei, D. E. Cox, and G. Shirane, "Magnetic Structures in the MnSb-CrSb System", Physical Review,  129, No. 5, pp. 2008--2018 (1963)

\bibitem{crsbin} V.D. Esin, D.Yu. Kazmin, Yu.S. Barash, A.V. Timonina, N.N. Kolesnikov, and E.V. Deviatov, "Josephson diode and spin-valve effects on the surface of altermagnet CrSb", 	JETP Letters, 123, 556 (2025) DOI: 10.1134/S0021364026600667
\bibitem{mntein}  D.Yu. Kazmin, V.D. Esin, Yu.S. Barash, A.V. Timonina, N.N. Kolesnikov, E.V. Deviatov, "Andreev reflection for MnTe altermagnet candidate", 	Physica B: Condensed Matter, 696,  416602 (2025) 	https://doi.org/10.1016/j.physb.2024.416602
\bibitem{cdas} O. O. Shvetsov, V. D. Esin, A. V. Timonina, N. N. Kolesnikov, and E. V. Deviatov, "Surface superconductivity in a three-dimensional Cd3As2 semimetal at the interface with a gold contact" Phys. Rev. B 99, 125305 (2019)
\bibitem{cosns} O. O. Shvetsov, V. D. Esin, A. V. Timonina, N. N. Kolesnikov, E. V. Deviatov, "Multiple magnon modes in the Co3Sn2S2 Weyl semimetal candidate",  EPL, 127, 57002 (2019)
\bibitem{timnal} V. D. Esin, D. N. Borisenko, A. V. Timonina, N. N. Kolesnikov, and E. V. Deviatov, "Spin-dependent transport through a Weyl semimetal surface" Phys. Rev. B 101, 155309 (2020)
\bibitem{black} N. N. Orlova, N. S. Ryshkov, A. A. Zagitova, V. I. Kulakov, A. V. Timonina, D. N. Borisenko, N. N. Kolesnikov, and E. V. Deviatov, "Band gap reconstruction at the interface between black phosphorus and a gold electrode" Phys. Rev. B 101, 235316 (2020)

\bibitem{planarHall} Guoxin Zheng, Arjyama Bordoloi, Mingjun Fan, Shunsuke Kitou, Hiraku Saito, Taro Nakajima, Sobhit Singh, Takashi Kurumaji, and Linda Ye, "Dominant in-plane anomalous Hall effect in a monoclinic room-temperature ferromagnet", arxiv:2606.10063

\bibitem{bow-tie} Felipe Tejo, Denilson Toneto, Sim\'on Oyarz\'un, Jos\'e Hermosilla, Caroline S. Danna, Juan L. Palma, Ricardo B. da Silva, Lucio S. Dorneles, and Juliano C. Denardin, "Stabilization of magnetic skyrmions on arrays of self-assembled hexagonal nanodomes for magnetic recording applications" ACS Appl. Mater. Interfaces, 12, 47, 53454 (2020).
\bibitem{cosnsmag} A.A. Avakyants, N.N. Orlova, A.V. Timonina, N.N. Kolesnikov, E.V. Deviatov, "Evidence for surface spin structures from first order reversal curves in Co3Sn2S2 and Fe3GeTe2 magnetic topological semimetals", 	Journal of Magnetism and Magnetic Materials, 573, 170668 (2023), 	https://doi.org/10.1016/j.jmmm.2023.170668.

\bibitem{ahe5} Tianye Yu, Ijaz Shahid, Peitao Liu, Ding-Fu Shao, Xing-Qiu Chen and Yan Sun, "Néel vector-dependent anomalous transport in altermagnetic metal CrSb" npj Quantum Materials 10, 47 (2025). https://doi.org/10.1038/s41535-025-00766-3

\bibitem{crsb_magnetization} N.N. Orlova, A.A. Avakyants, V.D. Esin, A.V. Timonina, N.N. Kolesnikov, E.V. Deviatov, "Altermagnetic  bulk and topological surface magnetizations for CrSb single crystals", 	arXiv:2512.11344

\bibitem{golub_alter} L. E. Golub, L. Šmejkal "Spin orientation by electric current in altermagnets", 	arXiv:2503.12203

\bibitem{2wMagF1} Debottam Mandal, Kamal Das, Amit Agarwal, "Chiral anomaly and nonlinear magnetotransport in time reversal symmetric Weyl semimetals"	Phys. Rev. B 106, 035423 (2022).
\bibitem{2wMagF2} A. A. Zyuzin and A. Yu. Zyuzin, "Chiral anomaly and second-harmonic generation in Weyl semimetals" Phys. Rev. B 95, 085127 (2017). DOI: 10.1103/PhysRevB.95.085127 

\end{thebibliography}
\end{document}